\documentclass{aa}  

\usepackage{graphicx}
\usepackage{txfonts}
\newcommand{\degree}{\ensuremath{^\circ}}
\begin{document}

%   \title{ALMA reveals a rotating [CII] disk in a gas-rich starburst galaxy \\at $z$=4.8}
%   \title{ALMA reveals a nascent galaxy at $z$=4.8\\ embedded in a turbulent rotating [CII] disk}
   \title{ALMA resolves turbulent, rotating [CII] emission\\ in a young starburst galaxy at $z$=4.8}

   \author{Carlos De Breuck\inst{1}
          \and
          Rebecca J. Williams\inst{2}
          \and
          Mark Swinbank\inst{3}
          \and
          Paola Caselli\inst{4}
          \and
          Kristen Coppin\inst{5}
          \and
          Timothy A. Davis\inst{1}
          \and
          Roberto Maiolino\inst{2}
          \and
          Tohru Nagao\inst{6}
          \and
          Ian Smail\inst{3}
          \and
          Fabian Walter\inst{7}
          \and
          Axel Wei\ss\inst{8}
          \and
          Martin A. Zwaan\inst{1}
          }

\institute{European Southern Observatory, Karl Schwarzschild Stra\ss e 2, 85748 Garching, Germany
\and
Cavendish Laboratory, University of Cambridge, 19 J. J. Thomson Avenue, Cambridge CB3 0HE, United Kingdom
\and
Institute for Computational Cosmology, Durham University, South Road, Durham DH1 3LE, United Kingdom
\and
School of Physics and Astronomy, University of Leeds, Leeds LS2 9JT, United Kingdom
\and
Centre for Astrophysics, Science \& Technology Research Institute, University of Hertfordshire, Hatfield AL10 9AB, United Kingdom
\and
Research Center for Space and Cosmic Evolution, Ehime University, Bunkyo-cho 2-5, Matsuyama, Ehime 790-8577, Japan
\and
Max-Planck Institut f\"ur Astronomie, K\"onigstuhl 17, D-69117, Heidelberg, Germany
\and
Max-Planck-Institut f\"ur Radioastronomie, Auf dem H\"ugel 69, D-53121 Bonn, Germany
             }

   \date{Received 2013 December 23; accepted 2014 April 3}

% \abstract{}{}{}{}{} 
% 5 {} token are mandatory
 
   \abstract {We present spatially resolved Atacama Large
       Millimeter/submillimeter Array (ALMA) [C{\sc ii}] observations
     of the $z=$\,4.7555 submillimetre galaxy, ALESS\,73.1.  Our
     0\farcs5 {\sc fwhm} map resolves the [C{\sc ii}] emitting gas
     which is centred close to the active galactic nucleus
     (AGN). The gas kinematics are dominated by rotation but with high
     turbulence, $v_{\rm rot}/\sigma_{\rm int}\sim$\,3.1, and a Toomre
     $Q$ parameter $<$1 throughout the disk. By fitting three
       independent thin rotating disk models to our data, we derive a
     total dynamical mass of
     3$\pm$2\,$\times$\,10$^{10}$\,$M_\odot$. This is close to the
     molecular gas mass derived from previous CO(2-1) observations,
     and implies a CO to H$_2$ conversion factor $\alpha_{\rm
         CO}$$<$2.3\,M$_{\odot}$(K\,km\,s$^{-1}$\,pc$^2$)$^{-1}$.  The
       mass budget also constrains the stellar mass to
       $<$3.1$\times$\,10$^{10}$\,$M_\odot$, and entails a gas
       fraction of $f_{\rm gas}\gtrsim$\,0.4. The diameter of the dust
       continuum emission is $<$2\,kpc, while the star-formation rate
       is as high as 1000\,M$_{\odot}$yr$^{-1}$. Combined with our
       stellar mass constraint, this implies an extreme specific star
       formation rate $>$80\,Gyr$^{-1}$, especially since there are no
       clear indications of recent merger activity.  Finally, our
     high signal-to-noise [C{\sc ii}] measurement revises the observed
     [N{\sc ii}]/[C{\sc ii}] ratio, which suggests a close to
     solar metallicity, unless the [C{\sc ii}] flux contains
       significant contributions from H{\sc ii} regions. Our
     observations suggest that ALESS73.1 is a nascent galaxy undergoing
     its first major burst of star formation, embedded within an
     unstable but metal-rich gas disk. }


   \maketitle
%
%________________________________________________________________

%
%
%
\section{Introduction}

The [C{\sc ii}]\,$\lambda$157.74\,$\mu$m line is a powerful alternative line
to $^{12}$CO for studying the interstellar medium in high-redshift
galaxies \citep[e.g. review by][]{carilli13b}.  The [C{\sc ii}] line arises
predominantly from photodissociation regions (PDRs) associated with
star-forming regions; other contributions come from diffuse H{\sc i}
clouds, low-density warm gas, or denser H{\sc ii} regions
\citep[e.g.][]{madden97}, and possibly from shock enhancement
\citep[e.g.][]{appleton13}.  In the most active systems, [C{\sc ii}] is the
dominant cooling line, representing $\sim $\,0.1--1\% of the total
luminosity \citep[e.g.][]{stacey91}.  This luminosity means that [C{\sc ii}]
has significant promise as a route for determining redshifts of even
the most obscured systems \citep{swinbank12b,weiss13}.

Until recently, [C{\sc ii}] has remained relatively unexplored in the local
Universe as its rest-frame wavelength requires balloon-borne or
space-based observations.  However, {\it Herschel}/PACS observations
have now begun to provide spatially resolved [C{\sc ii}] maps at scales of
0.1–-1\,kpc in nearby galaxies \citep[e.g.][]{beirao12,parkin13}.
These observations show significant variations in the line and
continuum ratios involving [C{\sc ii}] due to a range of physical processes
including changes in the ionisation mechanism, gradients in
metallicity or radiation field strengths -- hinting at the potential
diagnostic power of this line.

Ironically, [C{\sc ii}] observations from the ground are easier at
$z>$\,1 as the line is redshifted into the submillimetre atmospheric
windows.  Over the past decade, [C{\sc ii}] detections have been
reported in an increasing number of galaxies at $z=$\,1--2
\citep{hailey10,stacey10,ferkinhoff13}, as well as more distant
$z>$\,2--6 systems
\citep{maiolino05,maiolino09,iono06,wagg10,ivison10,debreuck11,cox11,swinbank12b,venemans12,riechers13,wang13a,rawle13,neri14}.
The first of these high-redshift [C{\sc ii}] detections were made in
powerful quasars. These observations seemed to confirm the trend seen
in local galaxies, where the most luminous far-infrared sources
($L_{\rm FIR} >$\,10$^{11}$\,L$_\odot$) have a ratio of [C{\sc ii}] to
far-infrared (FIR) luminosity $L_{\rm [C{II}]}$\,/\,$L_{\rm FIR}$ that
is lower by about an order of magnitude
\citep[e.g.][]{luhman98,dios-santos13}.  However, subsequent
observations of a larger sample of powerful far-infrared sources,
less-dominated by powerful active galactic nuclei (AGNs), revealed that
the many high-redshift sources show [C{\sc ii}] lines with similar
$L_{\rm [C{II}]}/L_{\rm FIR}$ ratios to those of nearby normal
galaxies \citep{stacey10,carilli13b}.

By comparing the $^{12}$CO, [C{\sc ii}] and far-infrared luminosities in a
sample of $z$\,=\,1--2 galaxies, \citet{stacey10} showed that
star-formation dominated systems have similar $L_{\rm
  [C{II}]}$\,/\,$L_{\rm FIR}$ to local (lower-luminosity) normal
galaxies, while AGN dominated systems have lower ratios, as seen in
local ultra-luminous infra-red galaxies (ULIRGs).  In terms of PDR
models \citep[e.g.][]{kaufman99}, both classes are interpreted as
having kpc-scale emitting regions, but the AGN-dominated sources appear
to have an order of magnitude more intense far-UV radiation fields.

Luminous, high-redshift star-forming galaxies (submillimetre galaxies;
SMGs) rather than quasar hosts are ideal targets to study the ISM of
distant, luminous galaxies free from the influence of AGN.  Although
some SMGs contain luminous AGNs, it is clear from deep X-ray studies
that in $\sim$\,85\% of SMGs, the AGN does not dominate the bolometric
luminosity \citep{alexander05,georgantopoulos11}.  Indeed for
the less luminous SMGs the X-ray emission potentially originates from
star-forming processes rather than an AGN \citep[e.g.][]{wang13b}.

In this paper, we present new ALMA observations which spatially
resolve the [C{\sc ii}] emission around an SMG at $z=$\,4.76 in the
Extended {\it Chandra} Deep Field South (ECDFS): ALESS\,73.1 (also
known as LESS\,J033229.4$-$275619 or XID\,403).  This galaxy was
originally identified as a compact, high-redshift AGN
\citep{vanzella06,fontanot07} and also a faint X-ray source from the
{\it Chandra} observations of ECDFS \citep{gilli11} and was then
detected as the most likely counterpart of a luminous submillimetre
source in the LABOCA survey of ECDFS by \citet{weiss09}.  Its
properties, including both $^{12}$CO, [C{\sc ii}] and [N{\sc ii}] gas
emission, were studied in a series of papers
\citep{coppin09,coppin10,biggs11,wardlow11,debreuck11,nagao12,gilli14}.
Subsequent higher-resolution ALMA continuum observations by
\citet{hodge13} provide an unambiguous identification of the
$z$\,=\,4.76 source as a luminous SMG.  Our new ALMA observations map
the spatial distribution and kinematics of the [C{\sc ii}] and
rest-frame far-infrared emission within this system on $\sim $\,kpc
scales, providing new insights into the structure of the most vigorous
starbursts seen in the SMG population.

Throughout the paper, we assume $H_0=$\,73\,km\,s$^{-1}$\,Mpc$^{-1}$,
$\Omega_m=$\,0.27, and $\Omega_{\Lambda}=$\,0.73, indicating a scale
size of 6.4\,kpc/\arcsec at $z=$\,4.76.

%__________________________________________________________________

%
%
%
\section{Observations}
\subsection{ALMA}
The ALMA observations of ALESS\,73.1 were obtained on UT 2012 July 18 and
27 in the compact Cycle~0 configuration, with 23 antennas covering
baselines between 18\,m and 402\,m.  We used the Band~7 receivers with
the four basebands centered at 317.20, 319.07, 329.54, and
331.11\,GHz. The [C{\sc ii}] line was covered in the last two basebands to
provide redundancy\footnote{As the real [C{\sc ii}] frequency turned out to
  be offset by 120\,MHz from the expected frequency predicted from the
  previous APEX observations of \citet{debreuck11}, the line was not well
  centred in the overlap region between the basebands. However, in
  hindsight, this overlap was not required, so this does not affect
  our data quality.}, while the other basebands covered continuum
emission.  The correlator was used in the frequency domain mode with a
bandwidth of 1875\,MHz (488.28\,kHz $\times$ 3840 channels per
baseband).  The total time on target was 1.62\,h.  The bandpass and
gain calibrators were J\,0522$-$364 and J\,0403$-$360, respectively;
the fluxes were calibrated with Pallas, Neptune, and Callisto.  The
observations were obtained in good weather conditions with
precipitable water vapour between 0.6 and 0.75\,mm.

\begin{figure}
   \centering
   \includegraphics[width=8.5cm,angle=0]{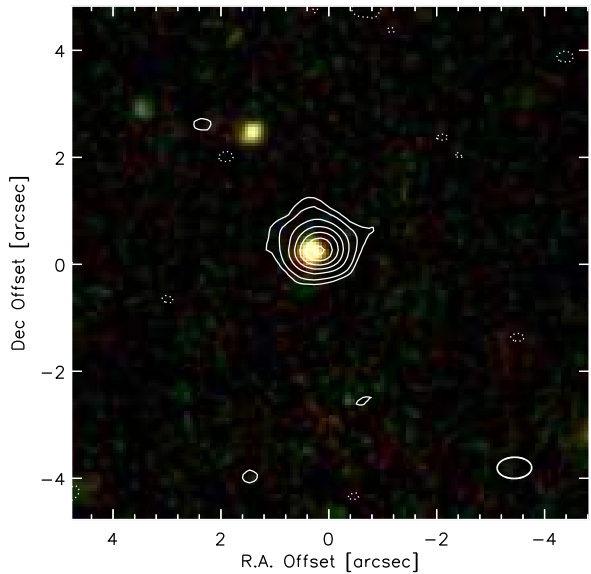}
   \caption{Continuum-subtracted, velocity-integrated [C{\sc ii}] emission
     (natural weighting) overplotted on a {\it HST} CANDELS $YJH$
     image \citep{grogin11,koekemoer11}.  The {\it HST} emission
     traced by is dominated by an unresolved point source, consistent
     with AGN-dominated rest-frame UV--optical emission, with no
     clear host galaxy visible.  Both the [C{\sc ii}], which is spatially
     resolved in our ALMA observations at $\sim$\,0\farcs6 resolution,
     and the spatially-unresolved continuum emission (see
     Fig.~\ref{channelmaps}) peak on the position of the
     optical/near-infrared source within 0\farcs2, indicating that the
     AGN lies close to the centre of the far-infrared emission.  The
     contour levels are $-$3, 3, 5, 10, 15, 20, 25, and 30\,$\sigma$, where
     $\sigma$=7.5\,mJy\,km\,s$^{-1}$. }
         \label{IJH_CII}
   \end{figure}

The data reduction followed the standard procedures in the Common
Astronomy Software Applications ({\sc casa}) package
\citep{petry12}. We applied a 22 channel (24.85\,km\,s$^{-1}$) binning
to the data cube, and cleaned it using natural weighting, which
provides a final synthesized beam size of
0\farcs65\,$\times$\,0\farcs40 at position angle of 91$^{\circ}$. To
search for spatially extended [C{\sc ii}] and continuum emission, we
also made maps with {\sc robust}\,=\,0.5, providing a resolution of
0\farcs56\,$\times$\,0\farcs38 (see Fig.~\ref{IJH_CII}).  In the
remainder of this paper, we only use the higher signal to noise ratio
(SNR) natural weighting data.  To derive continuum maps, we summed all
line free channels in the lower and upper sidebands separately,
resulting in two continuum maps at 318.13 and 330.34\,GHz, with
bandwidths of 3.78 and 3.44\,GHz, respectively (see
Table~\ref{parameters}). We also made a combined 318.13 + 330.34\,GHz
continuum map, which reaches an rms noise level of 0.15\,mJy.
% Table 1
\begin{table}
\caption{Observed properties of ALESS\,73.1}
\begin{tabular}{lll} 
\hline\hline
Parameter & Value & Reference 
\\  \hline  
$z_{\rm Ly\alpha}$     & 4.762\,$\pm$\,0.002 & 1 \\
$z_{\rm CO(2-1)}$      & 4.755\,$\pm$\,0.001 & 2 \\
$z_{\rm [CII]}$        & 4.7555\,$\pm$\,0.0001 & 3\\
$z_{\rm [NII]}$        & 4.7555\,$\pm$\,0.0002 & 4\\
$I_{\rm [CII]}$        & 7.42\,$\pm$\,0.35 Jy\,km\,s$^{-1}$ & 3\\
$L_{\rm [CII]}$        & 5.15\,$\pm$\,0.25\,$\times$\,10$^{9} L_{\odot}$ & 3\\
$L_{\rm 8-1000\mu m}$   & 5.6\,$\pm$\,1.5\,$\times$\,10$^{12} L_{\odot}$ & 5\\
$\Delta v_{\rm [CII]}$ & 161\,$\pm$\,45\,km\,s$^{-1}$ & 3 \\
$S(\rm 318.13\,GHz)$ & 5.9\,$\pm$0.1\,mJy & 3 \\
$S(\rm 330.34\,GHz)$ & 6.6\,$\pm$0.2\,mJy & 3 \\
$\rm [$C{\sc ii}] size\tablefootmark{a} & 0\farcs64 & 3 \\
Continuum size\tablefootmark{a} & 0\farcs29\,$\pm$\,0\farcs06  & 3 \\
$\rm [$C{\sc ii}] $\alpha$ (J2000) & 03:32:29.31 & 3 \\
$\rm [$C{\sc ii}] $\delta$ (J2000) & $-$27:56:19.66 & 3 \\
\hline
\end{tabular}
\tablebib{(1)~\citet{coppin09};
(2) \citet{coppin10}; (3) this paper; (4) \citet{nagao12};
(5) \citet{swinbank14}.}
\tablefoottext{a}{Deconvolved {\sc fwhm}}.
\label{parameters}
\end{table}

\subsection{APEX}
After the upgrade of the backend of the Swedish Heterodyne Facility
Instrument \citep[SHFI;][]{vassilev08} on APEX\footnote{APEX is a
  collaboration between the Max-Planck-Institut fur Radioastronomie,
  the European Southern Observatory, and the Onsala Space
  Observatory}, we re-observed the [C{\sc ii}] line in ALESS\,73.1 on UT
2013 July 10 to 13 (ESO programme 092.A-0668). The two 2.5-GHz wide
units of the Fast Fourier Transform Spectrometers now cover the full
4\,GHz IF bandwidth of SHFI, allowing a much improved baseline
subtraction compared to the original observations reported by
\citet{debreuck11}.  The data were obtained in good to excellent
observing conditions with a precipitable water vapour content of
0.16--1.05\,mm. The total integration time on source was 4.8\,h, and
the data were reduced using standard procedures in the Continuum and
Line Analysis Single-dish Software.

%
%                                                One column figure
%---------------------------------------------------CII over F125W
%
%______________________________________________________________

%
%
%
\section{Results}

%
%                                                One column figure
%----------------------------------------------------------spectra
   \begin{figure}
   \centering
   \includegraphics[width=8cm]{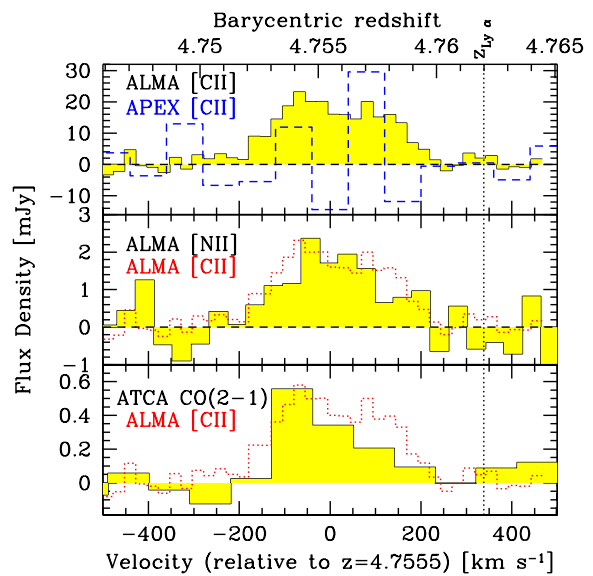}
   \caption{Comparison of the spatially integrated [C{\sc ii}], [N{\sc
       ii}], and $^{12}$CO(2--1) velocity profiles. A scaled version of
     the ALMA velocity profile is reproduced as a red dotted line in
     the middle and bottom panels to show that the other lines are
     consistent with it. The top panel shows the APEX [C{\sc ii}]
     non-detection (blue dashed histogram) is consistent within the
     uncertainties with the ALMA detection.  }
         \label{spectra}
   \end{figure}
%
%______________________________________________________________

\subsection{Integrated [C{\sc ii}] and continuum emission}
Table~\ref{parameters} lists the derived parameters from our ALMA
data. The velocity-integrated [C{\sc ii}] line is detected at a
significance level of 22$\sigma$ (Fig.~\ref{IJH_CII}), and is
spatially resolved with a deconvolved {\sc fwhm} size of
  0\farcs64 (4.1\,kpc) with the position angle unconstrained.  The
integrated line flux is only half the intensity originally reported by
\citet{debreuck11}.  Figure~\ref{spectra}, however, shows that the new
APEX observations with an rms=16\,mJy in 80\,km\,s$^{-1}$ channels do
not detect the [C{\sc ii}], which is consistent with the ALMA data.
The [C{\sc ii}], $^{12}$CO(2--1), and [N{\sc ii}] redshifts are now
all consistent within the uncertainties (see Table~\ref{parameters}),
suggesting they are originating from gas with the same bulk
  motion. We therefore adopt the [C{\sc ii}] redshift $z$=4.7555 as
the systemic (barycentric) redshift and zero-point of all
velocities quoted in this paper.

We also compared the velocity structure of the [C{\sc ii}] and the
[N{\sc ii}] data from \citet{nagao12} by making velocity slices along
the major axis of the data cube of both datasets.  The results are
consistent, though a detailed analysis is not possible due to the
limited SNR and spatial resolution of the [N{\sc ii}] data.  We
ascribe both the different [C{\sc ii}] flux level and the velocity
shift reported by \citet{nagao12} to baseline subtraction problems in
the 3\,$\times$ narrower bandwidth APEX spectroscopy of
\citet{debreuck11}.

The greyscales in Fig.~\ref{channelmaps} show the combined 318.13 +
330.34\,GHz continuum image, with a synthesised beam size of
0\farcs64\,$\times$\,0\farcs44. We very marginally spatially resolve
the continuum emission; the deconvolved {\sc fwhm} size is
0\farcs29\,$\pm$\,0\farcs06 (1.9\,$\pm$\,0.4\,kpc), with the position
angle unconstrained. To align the {\it HST} and ALMA images, we first
mapped the $BVR$ images from the MUltiwavelength Survey by Yale-Chile
\citep[MUSYC;][]{gawiser06} onto the ALMA astrometry from the ALESS
survey \citep[see \S2.2.1 of][]{simpson14}. We then translated this
new MUSYC astrometric solution to the {\it HST} images obtained as
part of the Cosmic Assembly Near-infrared Deep Extragalactic Legacy
Survey (CANDELS) images \citep{grogin11,koekemoer11}. The rms offset
between the {\it HST}, MUSYC, and ALMA astrometry is 0\farcs2, where
the uncertainty is mainly due to the use of galaxies rather than stars
for the alignment. We find that the ALESS73.1 dust continuum detection
is offset by 0\farcs22 from the {\it HST} F160W identification,
corresponding to $\sim$1$\sigma$ compared to the astrometric
accuracy. The {\it HST} image is itself spatially unresolved at
0\farcs20 (1.3\,kpc) resolution \citep{koekemoer11}. Hence both the
rest-frame UV and far-infrared emission point towards an apparently
compact galaxy. This is in contrast with the larger [C{\sc ii}] disk,
which has a projected {\sc fwhm} of 0\farcs68 (4.4\,kpc), and which
extends on both sides of the dust continuum (Fig.~\ref{channelmaps}).

We also use the continuum image to search for other sources within the
ALMA primary beam, we find no other detections above the
3\,$\sigma=$\,0.5\,mJy\,beam$^{-1}$ level. We can therefore
  exclude any moderately bright companions at the redshift of
  ALESS73.1 within a projected radius of 40\,kpc; scaling the
  star-formation rate (SFR) from the dust continuum flux of
  ALESS\,73.1 \citep{hodge13}, the non-detection implies that any
  companion should have an SFR of $<$\,75\,M$_{\odot}$\,yr$^{-1}$.
This indicates that ALESS\,73.1 is unlikely to be a component of a
mid-stage pre-coalescence merger with a second far-infrared-luminous
companion, unlike hyper-luminous sources like HATLAS\,J084933.4+021443
\citep{ivison13} or BR\,1202$-$0725
\citep{wagg12,carilli13a,carniani13}. A late stage merger cannot be
excluded with our current data.

\subsection{[C{\sc ii}] velocity analysis}
%
%                                                One column figure
%--------------------------------------CII channel maps over F160W
   \begin{figure*}
   \centering
   \includegraphics[width=13.5cm,angle=-90]{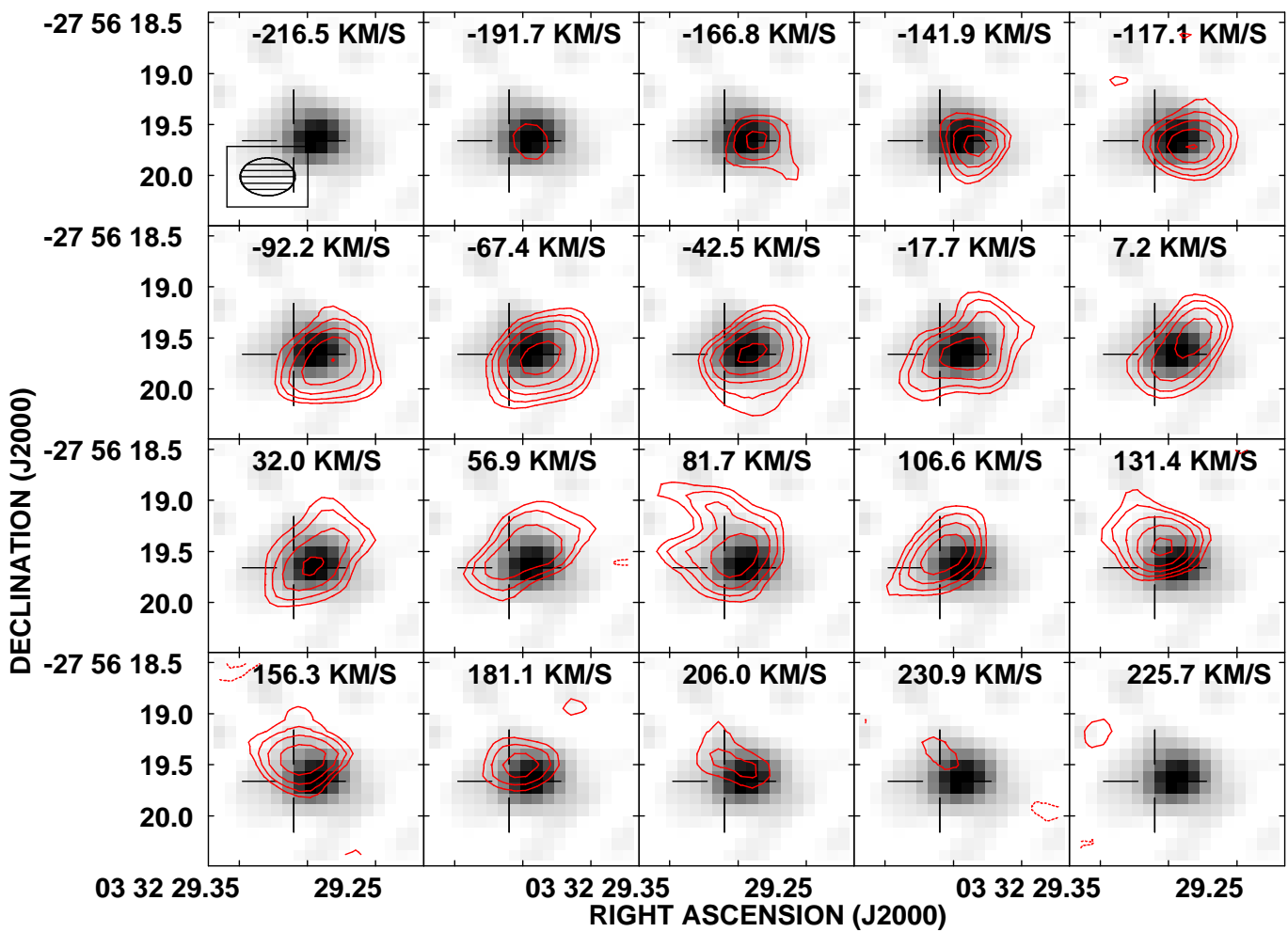}
   \caption{Channel maps (spacing 24.85\,km\,s$^{-1}$) of [C{\sc ii}]
     overlaid on the line-free 318.13+330.34\,GHz dust continuum
     image. Contour levels start at 3$\sigma$, where
     $\sigma$=0.7\,mJy/beam, and increase by $\sqrt 2$. The open cross
     marks the position of the unresolved F160W image
     (Fig.~\ref{IJH_CII}). We note that the dynamical centre coincides 
     with the compact dust continuum source. }
         \label{channelmaps}
   \end{figure*}

   Our high SNR ALMA observations spatially resolve the [C{\sc ii}]
   emission over 0\farcs6, i.e.\ spread over $\sim$\,2$\times$2
   synthesised beams (Fig.~\ref{channelmaps}). This is a significant
   improvement with respect to other ALMA and IRAM observations
   published to date, which have relatively low SNR and/or barely
   resolved the [C{\sc ii}] emission within the individual galaxies
   \citep{walter09,wagg12,gallerani12,carilli13a,riechers13,willott13,wang13a,carniani13,rawle13,neri14}. We
   note that several high redshift [C{\sc ii}] detections did detect
   nearby companion galaxies or different components of lensed
   galaxies.  Our ALMA data have a peak SNR=5--15 in each of the
   individual 25\,km\,s$^{-1}$ channels (Fig.~\ref{channelmaps}),
   allowing us to kinematically model the [C{\sc ii}] emission in this
   high redshift system, despite the rather limited spatial extent
   compared to the synthesised beam size \citep[see
   also][]{gnerucci11}. % for the first time.

\begin{figure*}
   \centering
   \includegraphics[width=18.5cm,angle=0]{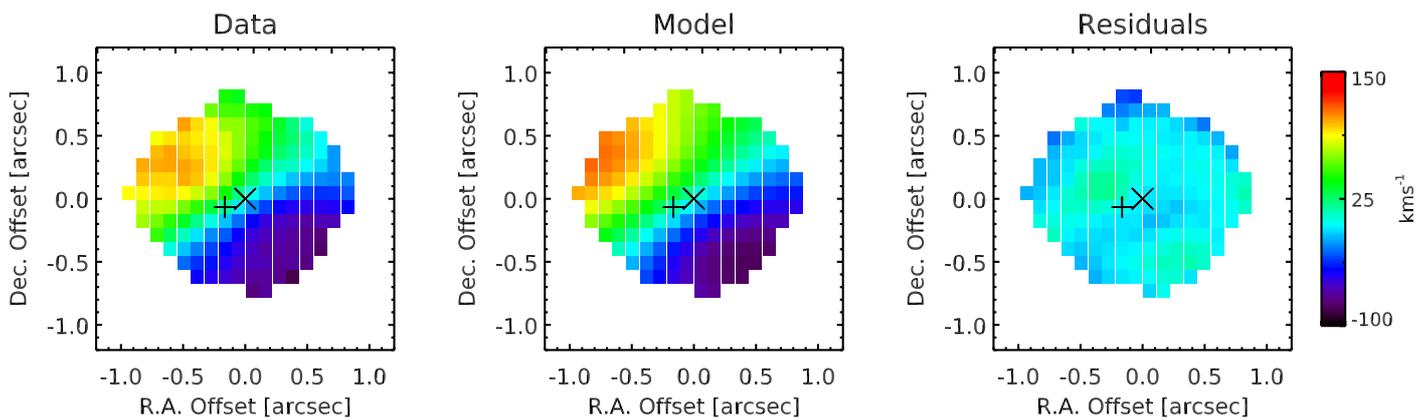}
   \caption{[C{\sc ii}] velocity field. The left panel presents the
     observed data, the central panel the best fit rotating disk model
     (see \S3.2) and the right panel the residuals. The plus and cross
     mark the locations of the optical counterpart
     (Fig.~\ref{IJH_CII}) and the [C{\sc ii}] peak flux (see
     Table~\ref{parameters}), respectively.  The observed motions of
     the [C{\sc ii}] emission are consistent with a rotating disk
     model.} \label{kinematics}
   \end{figure*}

\begin{figure*}
\includegraphics[width=6cm,angle=0]{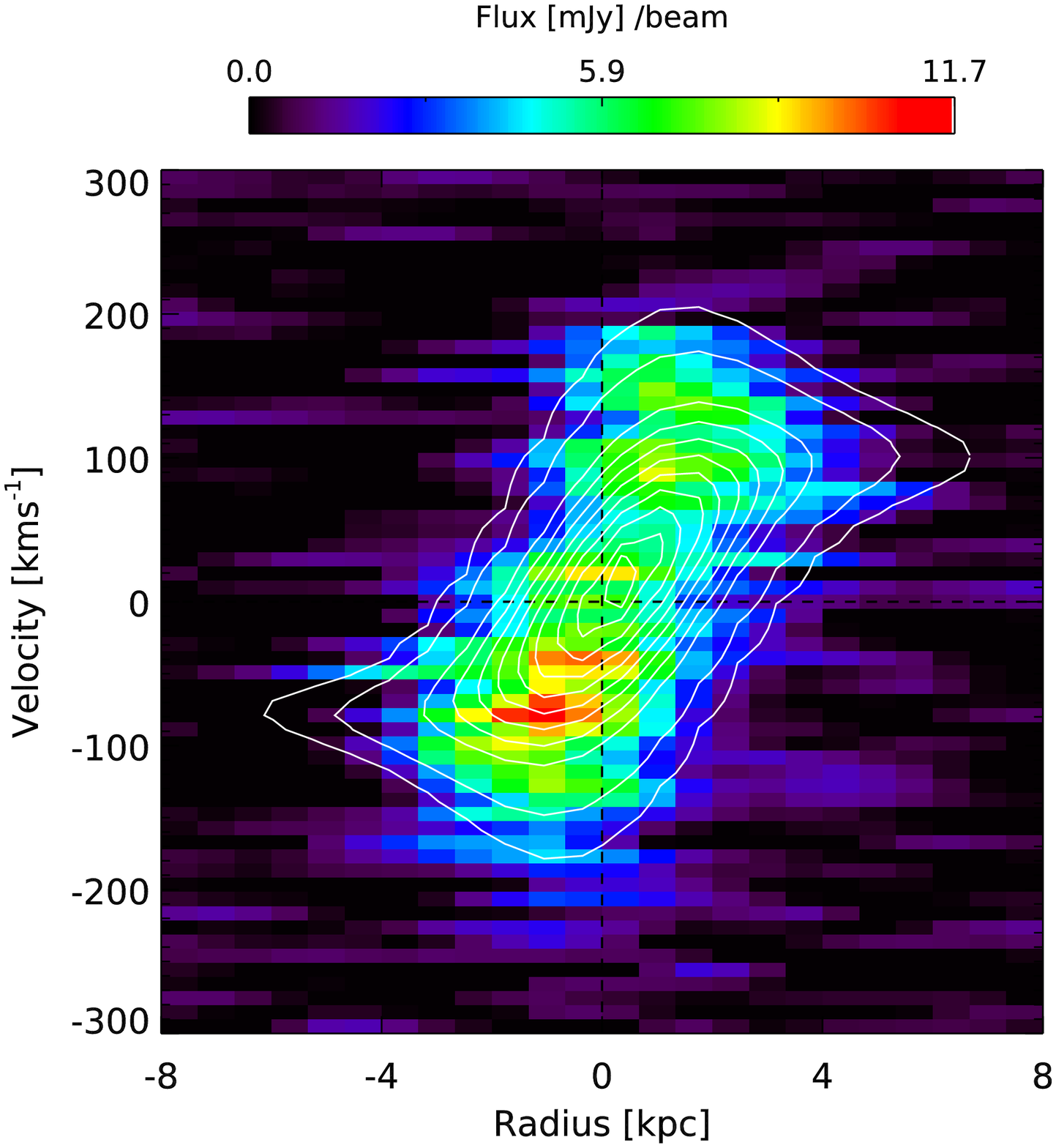}
\includegraphics[width=12cm,angle=0]{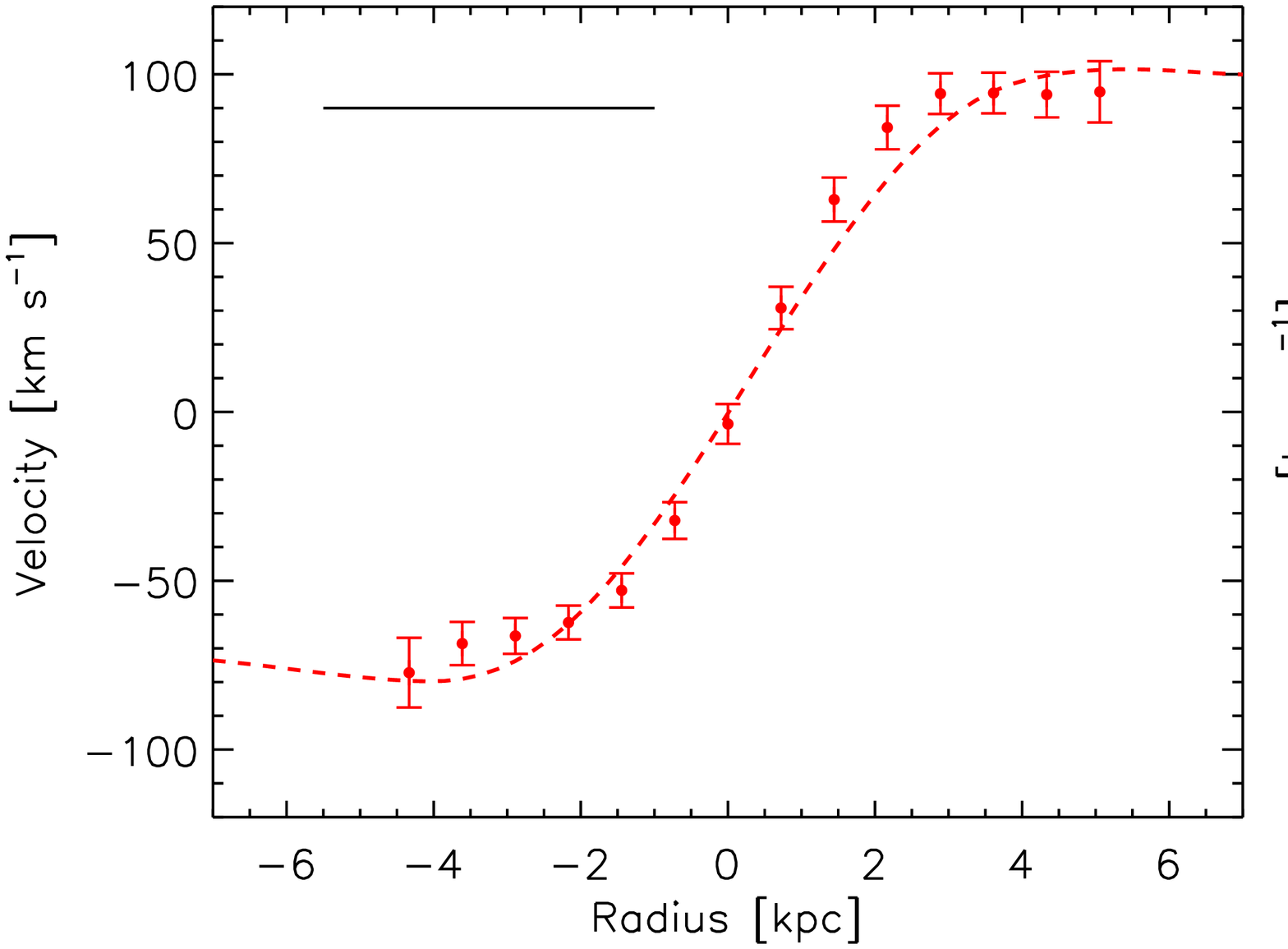}
\caption{({\it Left:}) Position-velocity diagram along the major axis
  of the disk model of Fig.~\ref{kinematics}. Contours show our best
  fit rotating disk model at 1\%, 5\%, 20\%, 30\%, 40\%, 50\%,
    60\%, 70\%, 80\%, 90\%, and 95\% of the peak flux.  ({\it Centre:})
  Rotation curve extracted along the major axis with our best-fit
  model overplotted as a dashed line. ({\it Right:}) The variation in
  velocity dispersion as a function of radius in the disk.  We
  estimate a typical $V_{\rm rot}/\sigma_{\rm int}\sim$\,3.1 in the
  disk, showing that it is relatively turbulent.  The rise towards the
  centre in the observed dispersion is an artifact resulting from the
  limited spatial resolution of the data. The synthesised beam size is
  shown as a horizontal black bar in the top left corner of the
  central and right panels. }
         \label{rotation_curve}
   \end{figure*}

The left panel of Fig.\ref{kinematics} shows the observed velocity
field of ALESS\,73.1 obtained by fitting the [C{\sc ii}] line with a single
Gaussian (results do not change significantly by fitting with two
Gaussians). The velocity field is dominated by rotation.  We
fit the velocity field with a dynamical model assuming that the
ionised gas is circularly rotating in a thin disk, and that the disk
surface mass density distribution is exponential $\Sigma(r)$=$\Sigma_0
e^{-r/r_0}$, where $r$ is the distance from the disk centre and $r_0$
is the scale radius. We neglect all hydrodynamical effects, therefore
the disk motion is entirely determined by the gravitational
potential. The model also includes the effect of beam smearing
\citep[for details, see][]{gnerucci10,gnerucci11}. The central panel
of Fig.~\ref{kinematics} shows our best fit (i.e. best-fit model
convolved with the beam). The right panel of Fig.~\ref{kinematics}
shows the residuals of the model, which are very small (less than 10
km/s in absolute value over most of the map). The bulk of the
velocity field is very well fitted by our simple rotating disk
  model, yielding  a maximum de-projected\footnote{Assuming
  $i$=53$^{\circ}$, see \S3.3.1.} velocity $v_{\rm
  rot}$=\,120$\pm$10\,km\,s$^{-1}$, oriented at a position angle
40$^{\circ}$$\pm$1$^{\circ}$ north through east
(Fig.~\ref{rotation_curve} {\it centre}). The effective
  half-light radius obtained by fitting a Gaussian to our model is
  2.4$\pm$0.2\,kpc; this is consistent with the 4.1\,kpc diameter {\sc
    fwhm} measured in the integrated [C{\sc ii}] image (see \S3.1), and with
  the turn-over radius of 2.2$_{-0.3}^{+2.0}$\,kpc obtained by
  modelling the shape of the rotation curve using the multi-parameter
  fit from \citet{courteau97}.
%______________________________________________________________
%
%                                                One column figure
%----------------------------------------------------------spectra

Fig.~\ref{rotation_curve} {\it left} shows a position velocity diagram
extracted from the cube across the major kinematic axis of the galaxy
with the model overplotted as contours. This highlights the
rotation seen in the system, but also shows that the luminosity
weighting of the [C{\sc ii}] is not constant; the brightest [C{\sc ii}]
originates from the higher velocity material.  Indeed, the ratio of
[C{\sc ii}] flux between $-$100 to $-$50\,km\,s$^{-1}$ and
50--100\,km\,s$^{-1}$ is 1.7 (see also Figs.~\ref{spectra} and
\ref{channelmaps}). This non-uniformity suggests that the disk is
either gas-loaded on one side, or preferentially illuminated on one
side; we return to this in \S3.4.

This flux asymmetry also illustrates the limits of our symmetric disk 
model. Significantly higher spatial resolution observations are needed
to determine a reliable flux distribution within the disk
  \citep[e.g.][]{deblok14}. In order to test the stability of
  our disk model and check if the non-uniform flux distribution in
the disk could bias our results, we also modelled the disk with
  two alternative models. First, we used the KINematic
Molecular Simulation ({\it KinMS}) routines of \cite{davis13} The
\textsc{KinMS} routine coupled to the Bayesian Monte Carlo Markov
Chain fitter {\it KinMS\_fit} (Davis et al., in prep.) matches the
brightness distribution of each pixel in the simulated and observed
datacubes, rather than fitting Gaussians like the fitting code
described above. Second, we used a simple arctan model
  \citep[e.g.][]{swinbank12a}, where the observed emission is fitted
assuming its rotation curve uses the form $v(r)=$\,2\,$\pi^{-1} v_{\rm
  asym} \arctan(r/r_t)$, where $v_{\rm asym}$ is the asymptotic
rotational velocity and $r_t$ is the effective radius at which the
rotation curve turns over. Both the alternative models, which
  have significantly different flux distributions from the model
  described above, obtain similar results. This provides confidence
  that our assumption of a rotating disk is a good (though not
  necessarily unique) representation of the observed [C{\sc ii}] velocity
  field. However, as we barely spatially resolve the flux distribution
  within the disk, we cannot distinguish which flux distribution is more
  appropriate. We will therefore quote the full range of uncertainties
  from all three models in any parameters derived from these models
  (notably the dynamical mass, see \S3.3.1).

  To derive the (model-independent) intrinsic velocity dispersion of
  the disk at each pixel (corrected for the contribution of the
  velocity gradient across the synthesised beam), we follow
  \citet{swinbank12a}.  At each pixel in the velocity dispersion map,
  we measure the luminosity weighted velocity gradient across the {\sc
    fwhm} of the beam at that pixel and subtract this from the
  velocity dispersion.  In Fig.~\ref{rotation_curve} {\it right}, we
  show both the observed and intrinsic one-dimensional velocity
  dispersion profile we derived, extracted along the major kinematic
  axis of the galaxy.  This shows that the intrinsic velocity
  dispersion of the disk is $\sigma_{\rm
    int}$\,=\,40\,$\pm$\,10\,km\,s$^{-1}$ (Fig.~\ref{rotation_curve}
  {\it right}). The ratio of rotational-to-dispersion-support 
    $v_{\rm rot}$\,/\,$\sigma_{\rm int}$\,=\,3.1\,$\pm$\, 1.0
  implying that this is a highly turbulent rotating disk. Such values
  are a factor of $\sim$three lower than local disk galaxies observed
  in CO \citep[e.g.][]{downes98}, but comparable to other
  high-redshift disks with similar resolution data from the H$\alpha$
  line
  \citep[e.g.][]{cresci09,genzel11,swinbank12a}. \citet{carniani13}
  reports a [C{\sc ii}] $v$\,/\,$\sigma$ $\sim$1.5 in both the SMG and
  quasar in the BRI1202 system, while in a quadruple system observed
  in CO, \citet{ivison13} report $v$\,/\,$\sigma$ $\sim$6 in the two
  brightest systems, and $v$\,/\,$\sigma$$<$1 in the faintest systems.
  We do warn that the optically thin [C{\sc ii}] emitting gas is not
  necessarily tracing the same gas phase as the optically thick low-J
  CO, so the higher intrinsic dispersion could also be due to the fact
  that we are observing a wider range of gas components.

%
%______________________________________________________________

%
%                                                One column figure
%----------------------------------------------------------spectra
%
%______________________________________________________________

\subsection{Constraints on dynamical and stellar mass}

\subsubsection{Dynamical mass}

\begin{figure}
\includegraphics[width=8cm,angle=0]{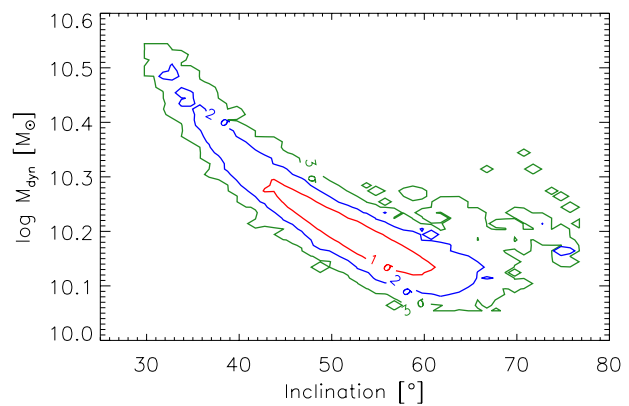}
\caption{Confidence contours for the dynamical mass and
  inclination of the rotating disk model from 10$^5$ trials of the
  MCMC.  We estimate a typical mass of the galaxy of $M_{\rm
    dyn}=$\,2.5$\pm$0.5\,$\times$\,10$^{10}$\,M$_{\odot}$
  within $R=$\,7\,kpc.  }
         \label{MCMC}
   \end{figure}

The circularly rotating disk model allows us to constrain the
dynamical mass $M_{\rm dyn}\sin^2(i)=R v_{\rm max}^2/G$, where
  $R$ is the radius at which the maximal velocity $v_{\rm max}$ is
  measured, and G the gravitational constant. To constrain the
  inclination, we use a Monte Carlo Markov Chain (MCMC) with 10$^5$
  realisations to investigate the uncertainties on the disk model
  parameters \citep[e.g.][]{carniani13}.  Fig.~\ref{MCMC} displays
the results of the MCMC in terms $M_{\rm dyn}$--$i$ confidence
contours.  The high quality of our data allows us to constrain the
inclination $i=$\,53\degree$\pm$9. This implies $M_{\rm
    dyn}$=\,1.4$\pm$0.5\,$\times$\,10$^{10}$\,M$_{\odot}$ within
  $R$=\,4\,kpc.  The alternative rotating disk models described in
  \S3.2 derive higher dynamical masses: the best fit \textsc{KinMS}
  model has a lower $i$=29\degree$\pm$4 and $M_{\rm
    dyn}$=$3.3_{-0.4}^{0.7}\times10^{10}$M$_{\odot}$, while the arctan
  model yields $i$=50\degree$\pm$8, and $M_{\rm
    dyn}$=4.1$\pm$0.9$\times10^{10}$M$_{\odot}$. Given the
  uncertainties in deriving an accurate flux distribution with our
  limited spatial resolution, we quote the full range of uncertainties
  from all three models, and adopt $M_{\rm
    dyn}$=\,3$\pm$2\,$\times$\,10$^{10}$\,M$_{\odot}$.

This dynamical mass is up to an order of magnitude
lower than those previously reported for $z$$\sim$2 SMGs on the
basis of resolved and unresolved $^{12}$CO kinematics for SMGs
\citep{tacconi08,swinbank11,bothwell13}.

\subsubsection{Stellar mass and AGN contributions}

The X-ray to radio spectral energy distribution (SED) of ALESS73.1
shows evidence of both AGN and stellar emission, with the bolometric
AGN contribution constrained to 2--20\%
\citep{coppin09,gilli14}. While the AGN dominates the mid-IR and
X-ray, the starburst component is dominating in the far-IR.  The
well-sampled rest-frame UV to near-infrared part of the SED has been
claimed to be consistent with a reddened stellar population
\citep[e.g.][]{coppin09,simpson14,gilli14,wiklind14}.  However, the
SED can be equally well fit by a single power law from the $U$-band to
24-$\mu$m in the observed frame, as expected from a reddened AGN
\citep{coppin09,wardlow11,simpson14}.  While the relatively weak
mid-IR emission implies that the AGN in ALESS73.1 contributes at
$<$20\% of the bolometric emission \citep{coppin09,gilli14}, the AGN
may still outshine the rest-frame UV/optical emission. Further
evidence of such a dominating UV/optical AGN contribution includes:
(i) The rest-frame UV spectrum presented by \citet{coppin09} which
displays narrow Ly$\alpha$ and $\sim$\,2000\,km\,s$^{-1}$ wide N{\sc
  v}\,$\lambda$1250\AA\ emission.  This spectrum is reminiscent of
that of the $z$\,=\,2.88 radio galaxy 4C\,24.28 \citep{rottgering97},
and is quite unlike the spectra of less-active star-forming galaxies
\citep[e.g.][]{shapley03}. A similar case to ALESS73.1 is
SMM\,J02399$-$0136 \citep{ivison98}, where the AGN light outshines the
rest-frame UV light from a vigorous starburst \citep{vernet01}.  (ii)
The unresolved morphology in the CANDELS {\it HST} images of
ALESS\,73.1 (Fig.\ref{IJH_CII}), indicating an upper limit on the size
of the UV source of $\leq $\,2\,kpc, which originally led to its
identification as a high-redshift AGN by \citet{fontanot07}. (iii) The
high X-ray luminosity of the source, whose spectrum has been claimed
to show a Compton-thick AGN \citep{gilli11}, and which is
  consistent with the observed N{\sc v} emission \citep{coppin09}.

  The stellar mass derived from an SED fit should therefore be
  considered as a strict upper limit. Assuming a standard
  light-to-mass ratio used by \citet{simpson14}, this implies
  $M_\ast$$<$\,7\,$\times$\,10$^{10}$\,M$_{\odot}$. As the unresolved
  {\it HST} F160W morphology suggests no significant host galaxy
  emission in addition to the bright AGN light, the actual stellar
  mass is most likely several times lower than this strict upper
  limit. Indeed, our limit on the dynamical mass, together with the
  gas mass of this system \citep[$M_{\rm
    H_2}$=1.6\,$\pm$\,0.3\,$\times
  $\,10$^{10}$M$_{\odot}$;][]{coppin10} imply
  $M_\ast$$\lesssim$\,3.1\,$\times$\,10$^{10}$\,M$_{\odot}$. We note
  that the $M_{\rm H_2}$ of \citet{coppin10} has been derived assuming
  a low $\alpha_{\rm CO}$=0.8\,M$_{\odot}$(K\,km\,s$^{-1}$\,pc$^2$)$^{-1}$
  conversion factor; adopting a higher $\alpha_{\rm CO}$ will decrease the
  limit on $M_\ast$. We return to this point in \S4.1.

  Alternatively, if the observed near-IR emission would be dominated
  by stellar light, the rest-frame $H$-band magnitude $M_{\rm
    H}$=$-$25.61 implies a light-to-mass ratio of $L_{\rm
    H}$\,/\,$M_\ast$$\gtrsim$\, 12. This is larger than the maximum
  light-to-mass ratio for all 77 ALESS SMGs, and 5$\times$ larger than
  the average \citep{simpson14}. We therefore exclude the possibility
  that the SED is dominated by stellar emission.

\subsection{Stability of the disk}

Not withstanding the variation in brightness within the disk, the
regular rotation pattern of the [C{\sc ii}] indicates that the gas it
traces had sufficient time to settle in a bulk motion, despite the
fact that the rotation period at the outer edge of the disk ($t_{\rm
  rot}\sim$\,200\,Myr) is 16\% of the age of the Universe at $z$=4.8.
At the same time, our compact and bright dust continuum detection
implies that the inner part of this disk is undergoing a violent burst
of star-formation with a rate of 1000\,M$_{\odot}$\,yr$^{-1}$
\citep[e.g.][]{coppin09,gilli14,swinbank14}. Such high SFR require a
high supply of cold gas. If the [C{\sc ii}] emitting gas is in a
regularly rotating pattern, could it still fuel the central starburst?
To answer this question, we now examine if this disk is dynamically
stable.

\citet{toomre64} introduced a stability criterion for thin rotating
disks against gravitational fragmentation. For gaseous disks
  \citep[e.g.][]{wang94}, this Toomre $Q$ parameter can be expressed
  as $Q(R) = {\sigma(R) \kappa(R)}/{\pi G \Sigma(R) > 1},$ where
$\sigma$ is the gas velocity dispersion, $\kappa$ the epicyclic
frequency and $\Sigma$ the gas mass surface density. All these
parameters depend on the radius $R$ from the centre of the rotating
disk. If the disk contains both gas and stars which gravitationally
interact with each other, the stability criterion becomes $Q=(1/Q_{\rm
  stars}+1/Q_{\rm gas})^{-1}>$\,1 \citep{wang94}. For ALESS\,73.1, we
cannot determine $Q_{\rm stars}$, but as argued above, we know that
the gas dominates the mass budget. Hence, if $Q_{\rm gas}$ is
unstable, the total system will be unstable.

Similar analysis have recently been carried out using spatially
resolved (adaptive optics assisted) IFU observations of the H$\alpha$
emission in $z\sim $\,1--2 galaxies, which have demonstrated that the
lowest $Q$ (most unstable) regions of high-redshift galaxies tend to
lie in the outer parts of the disks
\citep[e.g.][]{swinbank12a,genzel14}. However, H$\alpha$ traces the
ionised gas in the H{\sc ii} regions, which is an indirect tracer to
measure the properties of the star-forming gas, rather than the cold
molecular gas, which is a direct tracer of the fuel for star
formation. Although the [C{\sc ii}] line is tracing multiple gas
phases (see \S 1), it has been argued that this line is a good SFR
indicator
\citep[e.g.][]{stacey91,stacey10,delooze11,delooze14}. Detailed
observations of the [C{\sc ii}] kinematics are very difficult,
especially in the local Universe, as the line can only be observed
with small single-dish space or airborne telescopes. However, the
[C{\sc ii}], CO, and H{\sc i} velocity profiles do in general trace
each other rather well, although they can differ in some details
\citep{boreiko91,mookerjea11,braine12}. One example of a more detailed
study is the {\it Herschel} observations of the bright cluster galaxy
NGC4696, where the H$\alpha$ and [C{\sc ii}] emission trace each other
both morphologically and kinematically \citep[Fig. 6
of][]{mittal11}. Also at high-redshift, Gullberg et al. (in prep.)
found that the CO and [C{\sc ii}] velocity profiles in a sample of 13
gravitationally lensed submm galaxies are very similar. The most
likely alternative if the [C{\sc ii}] and CO are not tracing the same
bulk motion is that the lowest mass component (which in ALESS73.1 is
the atomic gas traced by [C{\sc ii}], see \S4.1) is outflowing
compared to the higher mass component. Although the SNR of the CO(2-1)
data is low, we do not detect any velocity shift (Fig.~\ref{spectra}),
nor do not find a significant outflow component in [C{\sc ii}] (see
\S4.2).  We therefore assume that the [C{\sc ii}] emission traces well
the kinematics of the underlying star-forming molecular gas component,
and derive the spatially resolved Toomre $Q$ in the molecular gas at
each pixel within the galaxy. Assuming a flat rotation curve
($\kappa$=$\sqrt{2} V/r$), and using the [C{\sc ii}] luminosity
distribution as a proxy to determine $\Sigma$, we calculate
$Q$($x,y$)\,=\,$\sqrt 2$\,$\sigma_{\rm int}(x,y) V(r)$\,/\,$\pi r G
\Sigma$ where $r$ is the radius from the dynamical centre of each
pixel, and $V(r)$ is the rotational velocity at radius $r$.

\begin{figure}
   \centering
   \includegraphics[width=8cm,angle=0]{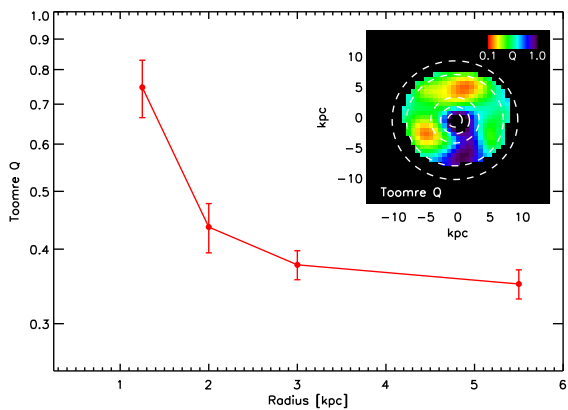}
   \caption{Spatial variation in Toomre $Q$ parameter within the gas
     disk in ALESS\,73.1. The main panel shows the azimuthally
     integrated $Q$ as marked with dashed lines in the inset.  We see
     that $Q$ declines with radius into the outer, gas-dominated,
     parts of the galaxy.  This behaviour is similar to that inferred
     for less actively star-forming galaxies at $z\sim$\,2 from
     H$\alpha$ kinematic studies
     \citep{genzel11,swinbank12a}. However, the entire disk has
     $Q$$<$1, meaning it is unstable throughout.}
         \label{fig:Qmap}
   \end{figure}

Fig.~\ref{fig:Qmap} shows the spatially resolved (beam smoothed)
Toomre $Q$ distribution, and an azimuthally averaged radial profile in
$\sim$\,1\,kpc bins across the galaxy image. At all radii, $Q$ is well
below 1, suggesting that the disk is unstable throughout. The average
$Q$ over the disk is 0.58$\pm$0.15, where the uncertainties include
the variation in inclination, weighting of the velocity gradient
correction, size of the disk, velocity field, and the removal size of
of the central aperture.
The increase in $Q$ towards the the inner regions is due to the high
torque on the gas making it more difficult to collapse, as often seen
in other high-redshift star-forming galaxies \citep[e.g.][]{genzel14}.

We conclude that the rotation of the [C{\sc ii}] emitting gas does not
prevent it from collapsing and being possible/likely fuel for the
violent starburst. Interestingly, the [C{\sc ii}] emission appears to
extend twice as far outwards than the dust continuum emission. This
could be due to a radial variation in the fraction of star-forming gas
traced by [C{\sc ii}]. Alternatively, this difference could just
reflect the mass distribution within the rotating disk, which our data
cannot reliably determine. Observations at higher spatial resolution
in [C{\sc ii}], and of more uniform $H_2$-tracers such as CO(1-0) or
[CI] are needed to obtain a more reliable distribution of the
star-forming gas.

%
%                                                One column figure
%----------------------------------------------------------spectra
%
%______________________________________________________________

%
%
%
\section{Discussion}
\subsection{A compact, gas rich galaxy with a high star-formation rate}

The $M_{\rm dyn}$=\,3$\pm$2\,$\times$\,10$^{10}$\,M$_{\odot}$
(\S3.3.1), is close to the cold gas mass $M_{\rm
  H_2}$=\,1.6\,$\pm$\,0.3\,$\times $\,10$^{10}$M$_{\odot}$, derived
from the $^{12}$CO(2--1) detection reported by \citet{coppin10}.  Note
that this $M_{\rm H_2}$ was derived assuming a conservative (low)
CO-to-H2 conversion factor $\alpha_{\rm
  CO}$=0.8\,M$_{\odot}$(K\,km\,s$^{-1}$\,pc$^2$)$^{-1}$, so the actual
$M_{\rm H_2}$ may still be significantly higher. It is therefore clear
that ALESS\,73.1 is a very gas-rich galaxy.  We can also obtain an
estimate of the atomic gas mass M$_{\rm a}$ associated with the
photodissociation regions using equation 1 from \citet{hailey10}.
We also note that this assumes that the fraction of the H$_2$ molecular gas
which not traced by CO is negligible. Such an assumption is
appropriate for ALESS73.1 as it has a moderately strong far-UV
ionisation field $G_0$$\sim$10$^3$ \citep{debreuck11} and close to
solar metallicity (see \S4.3). Following \citet{debreuck11}, we adopt
a C$^+$ abundance 1.4$\times$10$^{-4}$, a critical density $n_{\rm
  crit}$=2.7$\times$10$^3$\,cm$^{-3}$ and a PDR surface temperature
$\sim$300\,K. Using our more reliable ALMA [C{\sc ii}] flux
(Table~\ref{parameters}), we derive M$_{\rm a}$$\simeq
$\,4.7$\pm$0.5\,$\times $\,10$^9$\,M$_{\odot}$.  The combined atomic
plus molecular gas mass is therefore $\sim$\,2.1\,$\times
$\,10$^{10}$\,M$_{\odot}$, implying a gas fraction $f_{\rm
  gas}$=0.4--1.

We can also use the mass budget $M_{\rm H2}$$<$$M_{\rm dyn}-M_{\rm
  a}-M_\ast$ to constrain $\alpha_{\rm CO}$. Assuming no significant dark
matter contribution and minimising $M_\ast$, this puts an upper limit
$\alpha_{\rm CO}$$<$2.3\,M$_{\odot}$(K\,km\,s$^{-1}$\,pc$^2$)$^{-1}$, i.e. $<$ half of the Galactic value, but consistent with the range of $\alpha_{\rm
  CO}$ values found at high redshift
\citep[e.g.][]{ivison11,bothwell13,bolatto13}.

Despite being a relatively low-mass galaxy compared to dynamical
masses for SMGs \citep{tacconi08,bothwell13}, or masses estimated from
photometric modelling of the SEDs of submillimetre-selected galaxies
\citep[e.g.][]{hainline11,simpson14}, our bright and unresolved dust
continuum detection (Fig.~\ref{IJH_CII}) suggests that the
star-formation in ALESS\,73.1 is coincident with the position of the
AGN host galaxy, and not in a nearby companion, as seen in several
other SMGs \citep[e.g.][]{ivison08,ivison12,hodge13}. Using the
  {\it Herschel} 70--500\,$\mu$m limits, and three ALMA
  872--1305\,$\mu$m detections, \citet{gilli14} obtain $L_{\rm
    8-1000\,\mu m}$=\,5.9$\pm 0.9\times$\,10$^{12}$\,L$_{\odot}$. Any
  contributions to $L_{\rm 8-1000\,\mu m}$ from an AGN are constrained
  to 2--20\% thanks to the sensitive {\it Herschel} limits
  \citep{coppin09,gilli14}.  Using almost the same data, but adding
  also the the 20\,cm detection of \citet{miller13},
  \citet{swinbank14} find a very similar, $L_{\rm 8-1000\,\mu
    m}$=\,5.6$^{+1.8}_{-1.1}\times$\,10$^{12}$\,L$_{\odot}$, showing
  that the AGN contribution in also negligible in the radio.
Assuming the \citet{kennicutt98} relation, this $L_{\rm 8-1000\,\mu
  m}$ implies a SFR of 1000\,$\pm$\,150\,M$_{\odot}$\,yr$^{-1}$ for a
Salpeter initial mass function. This SFR can be compared to the empirical relations between SFR and the [CII] and [NII] luminosities derived by \citet{delooze14} and \citet{zhao13}. Using these, we find SFR=450$\pm$70 and 600$_{-500}^{+2700}$\,M$_{\odot}$\,yr$^{-1}$, respectively. This suggests the dust continuum derived SFR may be overestimated by a factor of two, but given the uncertain calibration of the [CII] and [NII] derived SFR at high redshift, we will adopt SFR=1000\,M$_{\odot}$\,yr$^{-1}$ in the remainder of this paper.

Combining this high SFR with the
$M_\ast$$<$\,3.1\,$\times$\,10$^{10}$\,M$_{\odot}$ limit (\S3.3.3)
implies a specific star-formation rate sSFR\,$>$\,80\,Gyr$^{-1}$ which
is significantly higher than the bulk ``normal'' star-forming galaxies
at the same redshift, i.e.\ almost an order of magnitude above the
``main sequence'' at the same redshift
\citep[e.g.][]{gonzalez10,stark13}. This galaxy will thus double its
stellar mass in $\sim$12\,Myr; we are thus likely observing ALESS73.1
during its first major burst of star formation.  Generally, galaxies
with sSFR well above the Main Sequence are identified as ``starburst
galaxies'' in which star-formation is thought to occur in a violent
mode as a consequence of galaxy merging or strong interactions.
Nevertheless, the atomic and ionised gas in ALESS\,73.1, as traced by
[C{\sc ii}], still shows rotationally supported disk kinematics (as
seen in less active, normal main sequence galaxies) underlining the
rapidity with which gas can reach such configurations, even in the
most active systems.
%
%                                                One column figure
%------------------------------------------------------metallicity
%
%______________________________________________________________

\subsection{Limits on  outflowing gas}
The [C{\sc ii}] profile in Fig.~\ref{spectra} shows no obvious
indication of an underlying broad component indicative of an outflow
as seen in some other high-z systems mapped in [C{\sc ii}]
\citep{maiolino12}. In order to set an upper limit on this outflow, we
fitted the ALMA spectrum by forcing an additional broad component with
a {\sc fwhm}\,$>$\,500\,km\,s$^{-1}$. This additional component is not
statistically significant and is not required by the fit, but we can
use it to set an upper limit of $<$\,1.3\,Jy\,km\,s$^{-1}$ on the
presence of an outflow. Using the same assumptions as in \S4.1, this
translates into an upper limit on the atomic gas mass in an outflow of
$<$\,9\,$\times$\,10$^8$\,M$_{\odot}$. Following \citet{canodiaz12},
we approximate the outflow as gas with constant velocity uniformly
distributed within a sphere of radius $R$. As we cannot determine $R$,
we assume $R$$\sim$\,1\,kpc as in massive outflows observed in other
galaxies at both high redshift
\citep[e.g.][]{maiolino12,canodiaz12,weiss12,carilli13a}, and low
redshift
\citep[e.g.][]{feruglio10,feruglio13,sturm11,aalto12,cicone12,cicone14,veilleux13}. We
can then constrain the outflow rate $\dot M=3\,v\,M_{\rm
  outflow}/R_{\rm outflow}$ \citep{maiolino12} to be
$\lesssim$\,1400\,M$_{\odot}$\,yr$^{-1}$. This is a rather loose upper
limit, but it does show that any outflow must be comparable or
  less than the star-formation rate, i.e.\ even if an outflow is
present, it is unlikely to dominate the evolution of the gas
reservoir.

\subsection{Metallicity}
%
%                                                One column figure
%------------------------------------------------------metallicity
   \begin{figure}
   \centering
   \includegraphics[width=8cm,angle=-90]{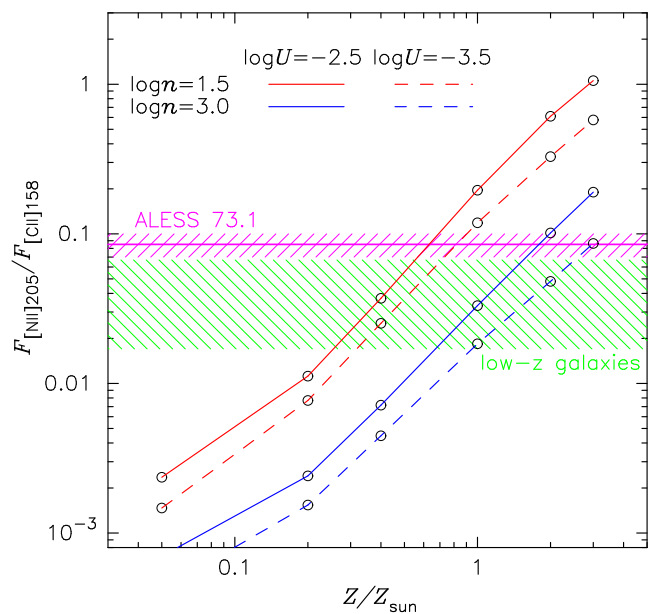}
   \caption{The observed [N{\sc ii}]/[C{\sc ii}] flux ratio compared
     with model predictions for different density $n$ and ionisation
     parameter $U$ \citep[see][]{nagao12}. The green hatched range
     denotes the observed range for low-redshift galaxies. The pink
     hatched range shows the observed ALMA ratio for ALESS\,73.1 with
     its uncertainty. The red and blue lines show {\sc Cloudy} model
     results as a function of $Z_{\rm gas}$ with $\log n_{\rm H[II]}
     =$\,1.5 and 3.0 respectively, while solid and dashed lines denote
     the models with $\log U_{\rm H[II]} -$2.5 and $-$3.5,
     respectively. We conclude that the metallicity in this gas rich
     disk is likely to be close to solar.}
         \label{metallicity}
   \end{figure}
%
%______________________________________________________________

   \citet{nagao12} used the [N{\sc ii}]\,205\,$\mu$m/[C{\sc
     ii}]\,158\,$\mu$m ratio in ALESS73.1 as a powerful new probe of
   the metallicity in the ISM. In particular, this line ratio is free
   of extinction effects which often complicate determinations from
   optical and near-infrared line ratios. The observed ratio used by
   \citet{nagao12} was rather uncertain due to the velocity offset
   seen with respect to the narrow-bandwidth APEX spectrum. Using the
   new [C{\sc ii}] spectroscopy, we can now constrain this ratio to
   [N{\sc ii}]/[C{\sc ii}]\,$=$\,0.085\,$\pm$\,0.015.
   Figure~\ref{metallicity} shows that this implies
   $Z=$\,0.6--3\,$Z_{\odot}$, where the uncertainty in dominated by
   the models rather than the measurement uncertainties. Our new ALMA
   [C{\sc ii}] data therefore strengthens the conclusion of
   \citet{nagao12} that ALESS\,73.1 already has gas with a
     metallicity close to solar, when the age of the Universe was a
   mere 1.2\,Gyr. Such highly enriched gas has been detected before in
   broad-line regions surrounding AGNs \citep[e.g.][]{hamann93}, but
   our ALMA observations now suggest that the highly enriched material
   may already be spread out over kpc scales.

   An alternative explanation for the relatively large [N{\sc
     ii}]/[C{\sc ii}] ratio is the "truncation" of the PDR. This has
   been reported by \citet{nakajima13} using optical observations of
   high-z Lyman$\alpha$ emitters.  Under the very strong radiation
   field due to the extreme star-formation, the relative volume ratio
   of H{\sc ii} regions and PDRs could change systematically in the
   sense that the relative PDR contribution becomes smaller for more
   active star-forming galaxies. This will reduce the [C{\sc ii}]
   flux, while the [N{\sc ii}] as a uniform H{\sc ii} tracer is not
   affected. The net result is then an increase in the [N{\sc
     ii}]/[C{\sc ii}] flux ratio for a given metallicity 
     \citep[e.g.][]{croxall12,decarli14}. As such, subsolar
   metallicities are still possible in ALESS73.1. Observations of
   other fine structure lines such as [N{\sc ii}]\,122$\mu$m, [O{\sc
     i}]\,144$\mu$m or [CI] are needed to determine the contributions
   from H{\sc ii} and PDRs, and hence derive a more accurate
   determination of the metallicity.

\section{Conclusions}

Using ALMA, we have spatially resolved the [C{\sc ii}]\,158\,$\mu$m
emission in the $z=$\,4.7555 SMG ALESS\,73.1.  The high SNR data cube
shows that the [C{\sc ii}] emitting gas extends twice as far out as
the dust continuum and exhibits with clear kinematical signatures of
rotation.  We demonstrate that these kinematical signatures are
well-described by a rotating disk with a disk size $\sim$\,2.4\,kpc
and a maximum deprojected rotation velocity of $v_{\rm
  rot}$=\,120\,$\pm$\,10\,km\,s$^{-1}$. The disk is highly turbulent
with a $v_{\rm rot}/\sigma_v\sim$\,3.1, and a Toomre $Q$ parameter
$<$1 throughout the disk, implying it is unstable.  Using this model
we constrain the dynamical mass of this galaxy to be 3$\pm$2\,$\times
$\,10$^{10}$\,M$_{\odot}$, which is close to the M(H$_2$) derived from
CO(2-1) \citep{coppin10}, and requires $\alpha_{\rm
  CO}$$<$2.3\,(K\,km\,s$^{-1}$\,pc$^2$)$^{-1}$. Combined with the
atomic mass M$_a$=4.7$\pm$0.5$\times$\,10$^{9}$\,M$_{\odot}$, our
dynamical mass constrains the stellar mass to
$<$3.1$\times$\,10$^{10}$\,M$_{\odot}$. Such low stellar mass is
remarkable for an isolated galaxy with a
SFR=1000\,M$_{\odot}$yr$^{-1}$, and suggests we are observing its
first major burst of star formation. Interestingly, our revised
integrated [N{\sc ii}]\,205\,$\mu$m/[C{\sc ii}]\,158\,$\mu$m ratio
suggest that the gas already has close to solar metallicity. However,
one should keep in mind that [C{\sc ii}] may have significant
contributions from non-PDR gas, which may have an impact on the
reliability of [N{\sc ii}]/[C{\sc ii}] as metallicity tracer and of
[C{\sc ii}] as a tracer of the star-forming gas
\citep{delooze11}. Spatially resolved observations of pure H$_2$
tracers such as the $J_{\rm upper}$$\leq$2 CO lines or the [CI] lines
\citep[e.g.][]{papadopoulos04} would be needed to provide a more
reliable determination of both the metallicity and the full extent of
the star-forming gas reservoir in ALESS73.1.

The ALMA Cycle 0 observations presented here, while still limited in
spatial resolution but not in SNR, illustrate the great potential of
ALMA to extend dynamical analysis tools out to the epoch when galaxies
had their first major burst of star-formation. Such observations of
atomic (or molecular) lines can be the only way to probe the dynamics
of even the highest redshift galaxies, which can be obscured or barely
resolved at optical/near-IR wavelengths. The final ALMA array will
increase its spatial resolution by an order of magnitude, allowing us to
regularly probe the scales of star-forming clouds, which have thus far
only been seen in studies of strongly lensed galaxies
\citep[e.g.][]{swinbank10}.

\begin{acknowledgements}

  We thank the anynymous referee for a very thorough reading of the
  manuscript, and numerous comments which have significantly improved
  the quality of the paper. We also thank Padelis Papadopoulos, Maud
  Galametz, Ilse De Looze, and Frank Israel for helpful suggestions,
  and the ALMA and APEX staff for their hard work behind the scenes to
  make these observations possible, and the ALMA Regional Centre staff
  to provide virtually science-ready data products.  IRS acknowledges
  support from STFC (ST/I001573/1), the ERC Advanced Investigator
  programme DUSTYGAL 321334 and a Royal Society/Wolfson Merit Award.
  AMS gratefully acknowledges an STFC Advanced Fellowship through
  grant number ST/H005234/1. The research leading to these results has
  received funding from the European Community's Seventh Framework
  Programme (/FP7/2007-2013/) under grant agreement No 229517. N is
  financially supported by JSPS (grant Nos. 23654068 and
  25707010). This paper makes use of the following ALMA data:
  ADS/JAO.ALMA\#2011.0.00124.S. ALMA is a partnership of ESO
  (representing its member states), NSF (USA) and NINS (Japan),
  together with NRC (Canada) and NSC and ASIAA (Taiwan), in
  cooperation with the Republic of Chile. The Joint ALMA Observatory
  is operated by ESO, AUI/NRAO and NAOJ. This work is based on
  observations taken by the CANDELS Multi-Cycle Treasury Program with
  the NASA/ESA {\it HST}, which is operated by the Association of
  Universities for Research in Astronomy, Inc., under NASA contract
  NAS5-26555.
\end{acknowledgements}

%-------------------------------------------------------------------

\end{document}